\documentstyle[11pt,epsf]{article}

\input psfig
\def\beq{\begin{eqnarray}}
\def\eeq{\end{eqnarray}}
\def\bea{\begin{eqnarray*}}
\def\eea{\end{eqnarray*}}

\def\calm{{\cal M}}

\def\centeron#1#2{{\setbox0=\hbox{#1}\setbox1=\hbox{#2}\ifdim
\wd1>\wd0\kern.5\wd1\kern-.5\wd0\fi
\copy0\kern-.5\wd0\kern-.5\wd1\copy1\ifdim\wd0>\wd1
\kern.5\wd0\kern-.5\wd1\fi}}
\def\ltap{\;\centeron{\raise.35ex\hbox{$<$}}{\lower.65ex\hbox{$\sim$}}\;}
\def\gtap{\;\centeron{\raise.35ex\hbox{$>$}}{\lower.65ex\hbox{$\sim$}}\;}

\def\singleandthirdspaced{\baselineskip=\normalbaselineskip\multiply
    \baselineskip by 130\divide\baselineskip by 100}
\def\singlespaced{\baselineskip=\normalbaselineskip}

\def\dslash{\not{\hbox{\kern-2pt $\partial$}}}
\def\Dslash{\not{\hbox{\kern-4pt $D$}}}
\def\Oslash{\not{\hbox{\kern-4pt $O$}}}
\def\Qslash{\not{\hbox{\kern-4pt $Q$}}}
\def\pslash{\not{\hbox{\kern-2.3pt $p$}}}
\def\kslash{\not{\hbox{\kern-2.3pt $k$}}}
\def\lslash{\not{\hbox{\kern-2.3pt $l$}}}
\def\qslash{\not{\hbox{\kern-2.3pt $q$}}}
\def\epsilonslash{\not{\hbox{\kern-2.3pt $\epsilon$}}}

\newcommand{\newc}{\newcommand}
\newc{\qbar}{{\overline q}}
\newc{\Kahler}{K\"ahler }
\newc{\deltaGS}{\delta_{\rm GS}}
\begin{document}
\begin{titlepage}
\begin{flushright}
{\large hep-th/0210256 \\ CALT-68-2409\\ RUNHETC-2002-36 \\  SCIPP-2002/21\\
}
\end{flushright}

\vskip 1.2cm

\begin{center}

{\LARGE\bf Supersymmetry, Axions and Cosmology}

\vskip 1.4cm

{\large T. Banks, M. Dine }
\vskip 0.2cm
{\it Santa Cruz Institute for Particle Physics,
     Santa Cruz CA 95064  } \\
\vskip 0.2cm
{\large M. Graesser} \\
\vskip 0.2cm
{\it California Institute of Technology, 452-48,
Pasadena, CA 91125}

\vskip 0.4cm

\vskip 4pt

\vskip 1.5cm

\begin{abstract}

Various authors have noted that in particular models, the upper
bound on the axion decay constant may not hold.  We point out that
within supersymmetry, this is a generic issue.  For large decay
constants, the cosmological problems associated with the axion's
scalar partner are far more severe than those of the axion.  We
survey a variety of models, both for the axion multiplet and for
cosmology, and find that in many cases where the cosmological
problems of the saxion are solved, the usual upper bound on the
axion is significantly relaxed.  We discuss, more generally, the
cosmological issues raised by the pseudoscalar members of moduli
multiplets, and find that they are potentially quite severe.

\end{abstract}

\end{center}

\vskip 1.0 cm

\end{titlepage}
\setcounter{footnote}{0} \setcounter{page}{2}
\setcounter{section}{0} \setcounter{subsection}{0}
\setcounter{subsubsection}{0}

\singleandthirdspaced


\section{Introduction}

There are three known solutions to the strong CP problem.   One is
the possibility that the $u$ quark mass vanishes.  While there are
arguments and lattice calculations which cast doubt on this
possibility, it is probably fair to say that it cannot be totally
ruled out within our present level of understanding.  The second
is that CP is a good symmetry of the underlying theory,
spontaneously broken in such a way that the observed $\theta$ is
very small (we will refer to this as the Nelson-Barr mechanism
\cite{nelson-barr}).
The third possibility is that there exists an axion.  The decay
constant (mass) of this axion is tightly constrained by
astrophysical and cosmological considerations.

>From the point of view of conventional effective field theory, the
axion idea seems at first sight implausible.   It requires
postulating a symmetry and then supposing that it is
only broken by tiny QCD effects.  Within the framework of
supersymmetric field theories, a sufficiently light axion to solve
the strong CP problem can be obtained with large discrete symmetry
groups.  For non-supersymmetric theories, the symmetries must be
{\it very} large.
But in string theory, axions which seem to have the
correct properties abound.  Indeed, the term ``axion" is used
rather generally for scalar fields with $2\pi$ periodicity (in
states which conserve CP, these fields are pseudoscalars).  Such
axions arise both in supersymmetric and non-supersymmetric string
backgrounds.  String theory axions are generally related to components
of antisymmetric tensor gauge fields, either in extra dimensions or the
usual four.  There are often arguments that in appropriate regions of
string theory moduli space, continuous shifts of these fields are
approximate symmetries which are broken primarily by the coupling to
low energy gauge multiplets.

If we assume low energy supersymmetry,
these axions are accompanied by (scalar) moduli.   With our current
understanding of string theory, we cannot determine
whether, in the full non-perturbative theory, any of these axions
can solve the strong CP problem, but it is quite plausible that they
do.  In this paper, we will distinguish generic axions by lower
case, and reserve ``Axion", with a capital A, for the QCD axion.

In this paper we will focus on two issues connected with axions
and their partners in supersymmetric theories.
\begin{itemize}
\item
With very mild assumptions, axions are light relative to their scalar
partners.   If the axion decay
constants are of order the Planck scale, the axions pose
cosmological problems significantly more serious than those
associated with scalar
moduli.  For smaller decay constants, these problems may be ameliorated.
We will discuss other possible solutions as well.
\item
The partner of the QCD Axion,
the Saxion, will have a mass of order the weak scale or smaller.  While
the existence of this particle has been noted in a number of
contexts, here we focus on a general point:  {\it
The cosmological problems posed by this modulus are far more severe
than those posed by the Axion.  So it does not make sense to discuss
cosmological limits on the Axion before considering a solution to
the Saxion problem.}
 Many solutions to the saxion problem
solve the Axion problem automatically.
In other words, {\it the usual cosmological upper bound
on the axion decay constant is not necessarily relevant to supersymmetric
theories}.
\end{itemize}

One of the principle suggestions to solve the cosmological moduli
problem\cite{moduliproblem,bkn} is that the (scalar) moduli are heavy, with
masses of order $10$'s of TeV, and that their decays reheat the
universe above nucleosynthesis temperatures.  One can object to this proposal
on the grounds that it is finely tuned, but we adopt it here
(it has been suggested that anomaly mediation might
offer an explanation of this surprisingly large mass scale\cite{randallmoroi};
two of us have argued elsewhere that anomaly mediation is not
generic in string/M theory\cite{sequestered}).  In this case, we
will see that there is a window in axion
decay constant, $f_a$,  around $10^{15}$ GeV \cite{liftingdecayconst}
in which
Axions might well be the dark matter.
A similar window, as we will see, exists for other possible
axions.

In the next section, we investigate the problem of axion mass.  We
focus on models in which supersymmetry is broken at an
intermediate scale, leaving low energy supersymmetry
breaking for Section 6.  We
explain why axions are parameterically light, and
discuss conditions under which there is an axion light enough to solve
the strong CP problem.  Section 3 contains a review of the QCD Axion
in the context of supersymmetry.  Some simple field theory models
which give variable $f_a$ are described.  In section 4, we review aspects of
 the conventional
moduli problem, and discuss the implications of light
pseudoscalars.  We enumerate possible
solutions of these problems.  In section 5, we turn to the cosmology
of the QCD Axion;
we explain why, for decay
constant less than $M_{Pl}$, the cosmological problems of the Saxion
are more severe than those of the Axion.  Possible solutions of
the Saxion problem, and their implications for the Axion, are
considered.  In general, we find that the Axion window extends up
to $f_a = 10^{15}$ GeV, but the precise limits are model-dependent\cite{liftingdecayconst}.
In section 6, we argue that such a large Axion window
the is probably not viable in
the context of gauge mediation, but noting that other solutions
are workable.
In section 7, we discuss alternative solutions to the
strong CP problem, and their possible implementations.

We also remark on some aspects of preheating in the context of
moduli.  While coherent production of scalars
during inflation has been discussed
as another aspect of the moduli problem, we explain why these are
essentially the same thing.  Coherent production of gravitinos has
also been discussed in the literature.  As we will see, the real
issue is production of ``modulinos."  This problem,
like the problem of axions,
is significantly less severe than the usual moduli problem.
Solutions to the latter generally resolve the former.
In our concluding section we discuss some remaining issues,
including some thoughts on axions in non-supersymmetric
models.  As this paper discusses many disparate issues, we present an overview in our
concluding section.

\section{Are Axions Light?}

In discussing the moduli problem in string
theory, one usually focuses on the ``scalar" component of the
moduli, and ignores the pseudoscalar.
The term pseudoscalar is, of course, a misnomer.  It implies an
unbroken CP symmetry.  The more precise distinction refers to the
fact that in string/M theory, in a modulus supermultiplet there is
typically a field with a $2 \pi$ periodicity.  The QCD Axion would
be one example of such a field.  We will follow a standard usage
in string theory, and refer to these fields
as axions, or occasionally simply pseudoscalars.  Both choices have their
linguistic limitations.  In general, we can ask how such fields can gain
mass, and how massive they might be.

The first question is whether the pseudoscalar is lighter than the
scalar field.  In principle, once supersymmetry is broken, both
fields can gain mass.  One might expect this mass to be comparable
to the mass of the scalar.  But in many pictures for how supersymmetry
might be broken, this is not the case.

String theory, if it describes nature, is presumably strongly
coupled.  On the other hand, it must also contain, effectively, small
parameters, if it is to explain the hierarchy
and the small values of the observed gauge
couplings.  We will
discuss two suggestions which have been put forward to
explain these facts shortly.
But independent of specific models, if one assumes that supersymmetry is related to the
solution of the hierarchy problem, one usually also assumes that the
superpotential, at the very least, is hierarchically small.  For example,
if the unified gauge couplings are related to a modulus, $S$, the
superpotential is assumed to be of order $e^{-S/b}$, for
some rational number $b$.

The periodicity properties of the
pseudoscalar can provide significant constraints on the
superpotential:
the
superpotential must take the form:
\beq
W = e^{-a {\cal M}} + e^{-b {\cal M}} + \dots
\eeq
for some constants $a$,$b,\dots, ~a<b<\dots$
So if we ignore possible symmetry
breaking from the Kahler potential, the full, supergravity potential will be of order
$e^{-2aM}$, but the leading terms which violate the Peccei-Quinn
symmetry
will be suppressed by a further exponential, $e^{-(b-a) \calm}$.

In making this argument, we are assuming that there are not large
corrections to the Kahler potential which violate the Peccei-Quinn
symmetry.  In principle, terms like
\beq
e^{-(\calm- \calm^\dagger)} f(\calm + \calm^\dagger)
\eeq
are consistent with the $2\pi$ shift symmetries.  It is usually
assumed that because such terms do not arise in string
perturbation theory, they are very small.  But we expect that the
minimum of the potential lies in a regime where perturbation
theory is not a useful guide.  For the superpotential, with
our assumption that $W$ is extremely small, holomorphy
still potentially provides significant constraints, but for the
Kahler potential, this is not the case.  Arguments have been given
that these terms might be small, but they are not based on
reliable calculations\cite{bdaxion}.  If these corrections are
${\cal O}(1)$, they would solve the problem of pseudoscalar moduli.  But
this would also mean that the axions seen in string perturbation
theory are not relevant to the solution of the strong CP problem.

We can examine proposals for supersymmetry
breaking in order to get some sense of what might happen.  Two suggestions for how moduli
might be stabilized while generating small gauge
couplings and large
hierarchies are known as ``Kahler stabilization"\cite{kahlerstabilization}
and the
``racetrack"\cite{racetrack}.

In Kahler stabilization, one assumes that
non-perturbative dynamics, such as gluino condensation,
generates a superpotential for the moduli.  For large values of the
moduli, the form of this superpotential can be determined.
For example, for the weak coupling dilaton, $S$, of the heterotic
string,
$W \sim e^{-aS}$, for some constant $a$.  In the Horava-Witten
limit, $W \sim e^{-aS-bT}$.
The
hypothesis is that the ground
state lies at large values of the moduli, so the superpotential
can be reliably calculated, but that the Kahler potential
cannot be calculated, and that it is the Kahler
potential which leads to stabilization of the moduli.
The motivation behind these proposals is that holomorphy constrains
the nonperturbative corrections to the superpotential to be exponential
in the moduli, while the Kahler potential is not similarly constrained.
In particular, in string perturbation theory we expect\cite{shenker}
corrections to the Kahler potential of order $e^{- b\sqrt{{\rm Re} S}}$.

In these models, the Kahler potential is also responsible for cancelling
the cosmological constant, so no field independent term is
added to the superpotential.
In the case of gaugino
condensation, assuming that there is only a single
modulus, $S$, for simplicity, the leading term in the superpotential has the form
$e^{-3S/b_o}$.  Requiring that this scale account for supersymmetry
breaking masses of order $1$ TeV, gives a scale of gaugino
condensation of order $10^{13}$ GeV.  This means
$e^{-S/b_o} \approx 10^{-5}$.

In the case of Kahler stabilization,
corrections to $W$ will arise, for example, from operators
such as $W_\alpha^2 W_\beta^2$ in the high scale
effective lagrangian.  These lead to a superpotential of
the form:
\beq
W = e^{- 3S/b_o} + e^{-6S/b_o}.
\eeq
This form is dictated by holomorphy and the non-anomalous
discrete $Z_2$ R-symmetry preserved by the higher-dimension
operator.
This superpotential
gives rise to a mass-squared for the pseudoscalar that is 
$e^{-3S/b_o}$
smaller than the mass squared 
of the scalar, and perhaps more importantly, a
mass
smaller than the scale of supersymmetry breaking by $e^{-3S/2b_o}
\sim 10^{-7.5}$.
So the pseudoscalar, in this picture, is much
lighter than the scalar.
Assuming, for example,
that the scalar moduli have masses of order TeV, the axion has a
mass of order $10$ KeV (if the mass is $100$ TeV, the axion has
mass of order $1$ MeV)

The pseudoscalar in this model is light, but it is not light
enough to play the role of the QCD Axion.  The QCD Axion might
simply lie in another multiplet.  This multiplet might not
couple to any gauge group stronger than QCD. Alternatively, there can be
a further suppression of the mass due to discrete symmetries.  A
discrete $Z_{2n}$
R symmetry can suppress $W_{\alpha}^{2n+2}$ type operators,
up to some value of $n$.  One needs roughly $n \ge 2$ to solve the
strong CP problem \footnote{The superpotential 
contribution to the Axion mass will not 
have its minimum at $\theta=0$. The bound comes 
from requiring 
that the full Axion potential which includes 
the usual 
QCD contribution, still has a minimum at $\theta \ltap 10^{-9}$.}.  
More generally, such symmetries could suppress
the masses of other axions, perhaps mitigating some of the
cosmological problems we will describe.

So far we have assumed there is
only one modulus and one strong group.
But there could be additional moduli that couple to other groups.
We assume, as before, that there is
no constant term in the superpotential. This is consistent
with the assumption that there is only one scale in string theory.
We also assume that for these moduli, supersymmetry breaking
results in a local minimum even if only the leading term in
the superpotential is retained. Further,
the dominant contribution to the moduli masses is assumed to come
from supersymmetry breaking.
Then the previous results still follow, namely, that
all the pseduoscalars will be
extremely light.



For the case of the racetrack, the situation is somewhat
different, and again model-dependent.  Indeed, there are not, to
our knowledge, entirely satisfactory models in which supersymmetry
is broken in this fashion, so we have to speculate even more about
what such a scheme might look like.  Roughly, however, we would
guess
that the superpotential will have the structure:
\beq
W= \alpha e^{-aS} + \beta e^{-bS}
\label{racetrackW}
\eeq
with $a$ and $b$ very nearly equal to each other.  One might
imagine
$a,b \sim {1 \over N}$, $a-b \sim {1 \over N^2}$, and $S = {1
\over g^2} \sim N^2$, for some large integer $N$.
If $\alpha $ and $\beta$ have the opposite sign there is a
locally supersymmetric minimum. Here the mass of the supermultiplet can
be large, unrelated to the size of supersymmetry breaking.
If  $\alpha $ and $\beta$ have the same sign there is no
local minimum. Brustein and de Alwis have argued for a
more general statement \cite{bd}. Assuming
a dilaton Kahler potential similar to that at tree
level, and that
the superpotential is steep, a characteristic of
racetrack models, they find that all local minima are also
locally
supersymmetric.

One may try to avoid this conclusion by adding in more fields:
\beq
W= X f(S) + Y g(S)
\eeq
where $f$ and $g$ have the racetrack form (\ref{racetrackW}),
but the parameters appearing in these functions are different.
For non-vanishing $S$ this model breaks supersymmetry.
Depending on the choice of parameters, local minima
can be found that break supersymmetry at large values of
$S$. These minima, however, are approximately
supersymmetric. For instance, in the limit 
that $ g << f$, the splitting within 
$S$ is set by $g$ and is tiny. 
By adjusting the parameters 
to increase $g \sim f$, one finds that the splitting
can be at most $O(1)$ before the
local minimum is lost.
If the (scalar and pseudoscalar) moduli
have masses sufficiently far above the scale of supersymmetry
breaking, there will be no moduli problem\cite{dineshirman}.


More generally, these last models, in the limit
that the supersymmetry-splittings are small,
are representative of a third possibility,
that some moduli are stabilized above the
scale of supersymmetry breaking, and that supersymmetry breaking
is a low energy effect, as in gauge mediation.  In this case,
there need be no moduli problem at all.  We will
comment on this possibility later.

A few remarks about the Axion multiplet in such models are in
order.  In general, there is no reason why the Axino and
Saxion masses shouldn't be of order $m_{3/2}$.  The operators:
\beq
\int d^4 \theta ({\cal A} +  {\cal A}^{\dagger})^2 Z^\dagger Z;
\int d^4 \theta ({\cal A} +  {\cal A}^{\dagger})^2 (Z+
Z^{\dagger}),
\eeq
where $F_Z \approx m_{3/2} M$, give mass to both fields.  One might
try to forbid the second operator by a symmetry, but such a
symmetry is also likely to forbid gaugino masses.

\section{The QCD Axion and Supersymmetry}

In this section, we consider some aspects of the QCD Axion in the
context of supersymmetry.  We review the general couplings of the
Axion and Saxion, and construct field theory models of axions with
various decay constants.  In effective field theory, we show
how, with suitable discrete symmetries, one can obtain an Axion
capable of solving the strong CP problem with a wide range of
decay constants.

\subsection{The QCD Axion and the Saxion}

As we have said, in general, the Axion must be part of a
superfield, ${\cal A}$, whose scalar component is $s+iA$.
$s$ is known as the saxion.
The existence of the axion coupling to QCD implies a
supersymmetric coupling:
\beq
L_{a}= {{\cal A}\over 16 \pi^2 f_a}W_{\alpha}^2.
\eeq
If there are gauge groups more strongly coupled than QCD, ${\cal
A}$ can couple to them only if they possess accidental
chiral symmetries which hold to a high degree of accuracy.
The field $s$ cannot gain mass larger
than some characteristic scale of supersymmetry breaking, since
otherwise the Axion would gain such a large mass as well.  The
axion multiplet also contains an axino, whose possible
cosmological implications have been widely discussed.  As we just
explained, one expects
that generally the axino is quite massive, so that the relevant cosmological
question is
the number of axinos produced subsequent to inflation.

\subsection{Models for Axions and Expectations for Axion Decay
Constants}

While string theory may be
the most robust context in which to consider axions, it is interesting
to consider other frameworks.  There are two reasons for this.
First, as we have noted, short distance, non-perturbative
effects in string
theory could spoil the Peccei-Quinn symmetry needed to solve the strong
CP problem.  Second, most frameworks for thinking about supersymmetry
in string theory constrain the Axion decay constant to be rather large,
typically of order the Planck scale, give
or take (possibly crucial) factors of $16
\pi^2$\cite{choi,bdaxion}. 
In the Horava-Witten limit of the heterotic string,
its decay constant can be still smaller\cite{bdaxion}.
Certainly numbers like $10^{15} ~{\rm GeV}$ don't seem at all
unreasonable.  But scales like $10^{11}$ GeV would seem more
likely associated with, say, the scale of supersymmetry breaking
in the effective low energy theory.

We can construct, as have many
authors\cite{dinelw,yamaguchi}, various field theory models which
give rise to accidental Peccei-Quinn symmetries.  One
can, in particular, construct models with superpotential:
\beq
W= c + {1 \over M_{Pl}^n}S^{n+2}S^\prime + S q \bar q + {1 \over M_{Pl}^{m+n}}
S^{m+n+2}S^\prime
+ \dots
\eeq
where $q$ and $\bar q$ carry color and perhaps weak
isospin and hypercharge; $c$ is a constant with dimensions of mass cubed,
$m$ is some large integer.
It is important that $m$ (and certain other integer
powers) be large in order that the Peccei-Quinn
symmetry (the approximate global symmetry which
rotates $S$ by a phase) hold to the required degree of approximation.
This structure can be enforced by discrete symmetries.
It is important, in this framework, that the symmetry not be an
$R$-symmetry, since this will be violated by the constant in $W$
required to cancel the cosmological constant.

Supersymmetry breaking will generate a potential for $S$ and
$S^{\prime}$ which can lead to a large expectation value for $S$.
Integrating out the fields $q$ and $\bar q$, leads to Axion-like
couplings for ${\cal A}= \ln(S)$.  In the case where supersymmetry
is broken in a hidden sector at an intermediate scale, $S$ and
$S^{\prime}$ will acquire susy-breaking masses of order $m_{3/2}$.
 With suitable signs for these terms,
the $S$ vev will be of order:
\beq
S = f_a \approx (m_{3/2} M_{Pl}^n)^{1 \over n+1}.
\eeq
The Saxion mass is of order $m_{3/2}$ for such models.
Note that for $n=1$, the Axion decay constant is about
$10^{11}$ GeV, in the range of decay constants in which Axions are
the dark matter according to the conventional analysis.

The Axion in this model gets a potential of the form
\beq
{\cal L}_{pq}=
{S^{2n+m+4} \over M_{Pl}^{2n+m}}
\eeq  (Other terms, such as those arising
from the $A$ terms, are of comparable size. Also,
the constant $c$ added to
cancel the cosmological constant leads to mass mixing
between the Axion and $S^{\prime}$, but this effect
is numerically
smaller than the term we describe here.)  This is suppressed
by $S^m \over M^m_{Pl}$ relative to the saxion mass squared.  While this is
a very small number, in order that
the QCD contributions dominate, it is necessary that $m$ be very
large, the precise value depending on the Axion decay constant
($<S>$).

As an aside,
note that this potential also illustrates the general issue we
discussed earlier of light pseudoscalars in moduli multiplets.
On the one hand, one requires a discrete symmetry to obtain
a large expectation value for $S$.  On
the other hand, this symmetry tends to make the pseudoscalar and
scalar masses very different.

An alternative possibility is that supersymmetry is broken at
lower energies, as in gauge-mediated models.  In this case, for
large $S$, one expects that the supersymmetry-breaking part of the
$S$ potential behaves as \cite{murayama}
\beq
V_s(S)= -\epsilon \vert F \vert^2 \left(\ln(\vert S \vert^2/M_{mess}^2)\right)^2.
\eeq
Here $\epsilon$ typically includes various loop factors, and so
may be rather small, and $M_{mess}$ is the messenger scale.
Assuming a superpotential of the form above, gives for $S$:
\beq
S \approx (\epsilon \vert F \vert^2 M_{Pl}^{2n})^{1 \over 2n+4}.
\eeq
The mass for the scalar is $m_s \sim \sqrt{\epsilon} F/S$.
But again, the pseudoscalar mass squared is suppressed by
the amount $S^m/M^m_{Pl}$.

\section{The Scalar and Pseudoscalar Moduli Problems}

\subsection{The Scalar Moduli Problem}

This problem is most concisely
phrased in the language of string theory (quantum gravity)\cite{bkn}.  One supposes that one
has some fields which vary over scale $M$, with a potential whose
characteristic size is $m_{3/2}^2 M^2$ ($M$ might be the Planck
or string scale, or, as we will discuss
later, $f_a$).  At early times, the characteristic
curvature of the modulus potential is of order $1/t \approx H$\cite{drt}.
So one expects the minimum, if any, of the modulus potential
to lie a distance of order $M$ from the flat space
minimum.   The modulus begins to oscillate about
its true minimum when $H \approx m_{3/2}$, at which time it
constitutes a fraction of order one of the total energy density of
the universe.  Even if the universe is radiation dominated at this
time, it quickly becomes matter dominated and remains so until the
modulus decays.  A plausible expectation for the decay width is
\beq
\Gamma = {1 \over 2\pi} {m_{3/2}^3 \over M^2}
\eeq
Assuming $m_{3/2} \approx 1 ~{\rm TeV}$ and $M \sim M_{Pl}$
this gives a decay time long after nucleosynthesis, and a
reheating temperature of order a few KeV.  This totally spoils the
successes of conventional big bang nucleosynthesis.

One possible solution to this problem is to suppose that the
minimum of the potential in the very early universe coincides with
the minimum now.  This is natural if the minimum preserves a
symmetry\cite{drt}.
Another
proposal to solve this problem is to assume that the modulus
is much more massive, with a mass more like $100$ TeV.  In this
case, the reheating temperature is about $7$ MeV, high enough to
restart nucleosynthesis.

Assuming that reheating is the solution, there are a number of issues
which one must address.  One is production of baryons\cite{bkn}.  The
dilution of any pre-existing baryons in these
decays is very substantial, a part in $10^7$.  In order that one
end up with a reasonable baryon density, one must suppose that
baryons are produced in the decays (through R-parity violating
operators), or that the pre-existing baryon density is very large
(e.g. as a result of very efficient AD Baryogenesis\cite{ad}).

Second, if there is a stable LSP, there is a potential issue with
overproduction of these objects\cite{lsps}.
But Moroi and Randall \cite{randallmoroi}
have pointed out that if the decaying
modulus is so heavy, the decays to LSP's can be suppressed by the
ratio of the LSP mass to the heavy
modulus mass (more generally the ratio of the masses of the MSSM
fields, squarks, sleptons, gauginos, etc.).  For example, couplings of
a modulus through kinetic terms to fermions, for on shell fermions,
are proportional to the mass of the fermion.  Similar, couplings to
scalars are proportional to the mass-squared of the scalars.
Decays to gravitinos
are potentially a problem, but one might expect that if the moduli
are so heavy, so are gravitinos, and that decays to gravitinos
might be kinematically forbidden.
Moroi and Randall indeed argued that such a large modulus (and gravitino) mass
might arise naturally in anomaly mediated supersymmetry breaking.

In thinking about axions, we will want to consider more general
possibilities for the scalar ``decay constant," i.e. the scale
over which the field varies, as well as its couplings to gauge
fields and other fields.  As the decay constant of the scalar
becomes smaller, adequate reheating can be obtained with a smaller
mass.  In particular, a mass of order 1 TeV and
decay constant of order $10^{15}$ GeV gives
reheating to $10$ MeV.  However, the lower mass raises again the
specter of overproduction of LSP's.  In this case, one might
require either breaking of R parity, or a wino LSP (see
\cite{randallmoroi}).

\subsection{Scalar Moduli Decays}

The decays of the saxion have been discussed in the literature,
and decay of the scalar moduli raise similar issues.  Scalars
can, in principle, decay to quarks and leptons, to gauge
fields, and to axions.  The decays of saxions to gauge bosons are
model-independent, and can be parameterized in terms of $f_a$.
The decays to axions and matter fields are model-dependent.  Among
familiar string moduli, at weak coupling the heterotic string field $S$ couples at tree
level to axions, but not to matter fields; the $T$ modulus couples both
to axions and to matter fields.  In
the class of field theory models described above, in which $S$ couples to the
vector-like pair $q,\bar q$, there are tree level couplings of the saxion to
axions, of the form
\beq
{h \over f_a} s (\partial_{\mu} a)^2
\eeq
where $h$ is a constant of order unity.  There are no tree level
couplings to quark and lepton fields.  Couplings to gauge
bosons arise at one loop.  As a result, the principle decay mode
of the saxion in such models is to axions.  As discussed
subsequently, this leads to cosmological difficulties.

It is possible to modify this class of models in such a way
that the saxion does couple to matter fields already at tree
level.  By suitably choosing discrete charges, one can arrange
that the Higgs boson transforms under the Peccei-Quinn
symmetry, giving rise to such tree level couplings, with
strength proportional to the Higgs Yukawa coupling.

There are two sets of issues involved with decays to axions.
First, in cases where the moduli don't couple to matter fields at
tree level, axions can easily end up carrying an order one
fraction of the energy when the scalars decay.  If the scalars
couple to matter fields, this fraction can be suppressed by the
number of matter fields.  However, if the axions are heavier than
roughly $1~{\rm MeV}$,
this may still be problematic, since they will come to dominate the
energy density before recombination \footnote{The actual bound on 
the axion mass from overclosure 
depends on the saxion mass, the second reheating temperature 
and the branching fraction decay of the saxion into axions. 
Using $(\rho_a/s)_0 \ltap \times 10^{-9} $ GeV, 
the constancy of $n_a/s$ and that $n_a = BF \times n_s$, 
one finds 
$m_a \ltap 10^{-7} m_s (\hbox{10 MeV})/(BF \times T^{(2)}_{RH})$. 
The bound quoted 
here is for $BF \sim 0.1$, $T^{(2)}_{RH} \sim 10$ MeV and 
$m_s \sim 1$ TeV.}.

\subsection{The Pseudoscalar Moduli Problem}

We
have seen that there are many possible values for the axion mass
and decay constant.  Consider, for example, the Kahler
stabilization model discussed above.  We indicated that for the
pseudoscalar component of ${\cal M}$, the mass is naturally
a few orders of magnitude below that of the scalar.  If
the decay constant of this axion is of order the Planck mass, then
when it begins to oscillate it carries an order one fraction of
the energy density of the universe.  Even if the decay of the
scalar reheats the universe to temperatures of order
nucleosynthesis temperatures, the axion quickly comes to dominate
the energy density, leading to an unacceptable cosmology.

If the axion decay constant is smaller, then the situation is
different.  First, the scalar modulus lifetime is much shorter, and for a
given scale of supersymmetry breaking, the reheating temperature
is higher.  For example, if the scalar mass is
$10^3$ GeV and the decay constant is $10^{15}$ GeV, the
reheating temperature is of order a few MeV.

Second, when the axion begins to oscillate, it carries a
fraction of the energy density of order $f_a^2/M_{Pl}^2$.  Now, if
the universe reheats to nucleosynthesis temperatures, this
fraction will be small enough if $f_a < 10^{15}$ GeV or so.  In
this case, at $10$ MeV, the energy fraction is less than
$10^{-6}$, so it is at most of order one at recombination.
So in this case,
both moduli problems are potentially solved.

We do, however, have to worry about the problem of scalar decays
to axions, discussed in the previous section.  As we have
indicated, in most models, the scalar already decays to axions at
tree level.
Scalar decay to axions
leads to two problems. During
nucleosynthesis axions increase the expansion rate
of the universe. This limits the branching fraction of saxion decays
into axions to be less than roughly 1/6, so that effectively
the axions do
not contribute more than one neutrino species.
This is easily satisfied in models
where the saxion also decays to a large number of standard
model species.


If there are additional axions, there are various possibilities
for their masses and couplings.  For example, if there is only one
strong gauge group, a second axion may play the role of the QCD
axion, provided that non-perturbative effects which break the PQ
symmetry are small enough.  A third axion might be much lighter
still, so that it does not pose significant cosmological
difficulties.

We can summarize, then, the solutions to the pseudoscalar moduli
problem:
\begin{itemize}
\item   The pseudoscalar is sufficiently heavy that its decays
restart nucleosynthesis.  This, however, requires that the scalar
moduli have masses of order $10^5$ TeV or larger.
\item  The scalar moduli masses are sufficiently large to restart
nucleosynthesis, while the decay constant is large
($f_a>10^{15}$ GeV),
and the pseudoscalars are extremely light, so
that they do not begin to oscillate until times of order
recombination time.  This requires that the pseudoscalar mass be
of order $10^{-36}(M_{Pl}/f_a)^4$ GeV.  This requires, in our discussion above,
very large discrete symmetries.
\item  There are order one {\bf PQ breaking} terms
in the Kahler potential, so the
pseudoscalar and scalar have comparable masses.  The main
difficulty with this proposal is that it leaves us without a
candidate for the QCD axion, except perhaps from something
generated at low energies as in our field theory
model above.
\item  The decay constant of the axion supermultiplet is $10^{15}$
GeV or smaller.  In this case, the scalar mass can be of order $1$
TeV, yielding sufficient reheating to restart nucleosynthesis.
The axion mass density is suppressed by $f_a^2/M_{Pl}^2$.
\item  As we will discuss later, another possibility is that there
are no axions.
\end{itemize}


\subsection{Other Cosmological Problems:  Axinos, Gravitinos,
Modulinos}

In the conventional picture the universe
just after inflation was described by a thermal
gas of particles and superparticles, and gravitinos
were produced in the collisions of these particles.
If these gravitinos
decayed during the era of nucleosynthesis, their abundance
must have been less than $n_{3/2} /n_{\gamma} \ltap 10^{-12}$
in order that not
too much helium was destroyed or
too much deuterium produced. Here we follow the convention 
of \cite{kawasakimoroi} that $n_{\gamma} \equiv \xi(3) T^3/\pi^2$ does not
include a factor of $g_*$, the number of relativistic
degrees of freedom.
This disruption to the production of 
light elements 
provides the strongest
constraint on the initial reheat
temperature of the universe \cite{ellis,kawasakimoroi}.
Just
after inflation, the gravitino to photon
ratio was roughly $n_{3/2}/n_{\gamma} \sim 
g_* \alpha T^{(1)}_{RH}/M_{Pl}
\sim T^{(1)}_{RH}/M_{Pl} $. But
since the number of
degrees of freedom has changed between the end
of inflation and the nucleosynthesis era, one
approximately accounts for this effect by
adding in
a dilution factor of $g_*(T^{(1)}_{RH})/g_*(keV)
\approx 100$. (This is just the statement that 
$n_{3/2}/s$ is conserved.) 
The 
gravitino abundance during the nucleosynthesis
era is properly obtained by
integrating the Boltzman equation, 
with the result depending on the gravitino mass.  
For $m_{3/2} \sim 1$ TeV 
the bound is approximately
\cite{ellis,kawasakimoroi}
\beq
\left({n_{3/2} \over n_{\gamma}}
\right)_0   \sim \left({ g_*(keV) \over g_*(T^{(1)}_{RH})} \right)
 {T^{(1)}_{RH} \over M_{Pl}} \sim 10^{-2} {T^{(1)}_{RH} \over M_{Pl}}
~.
\label{n32}
\eeq
This
implies a limit of roughly $10^{9}$ GeV
on the reheat temperature. Gravitinos heavier 
than about 3 TeV decay before nucleosynthesis, 
but can overproduce neutralino LSP's. In that case, 
the limit can be higher, up to $10^{12}$ GeV \cite{kawasakimoroi}.

The
late decay of a Saxion or saxion produces entropy,
which dilutes
any prexisting gravitinos or other relics.
The dilution factor is obtained in one of two equivalent
ways. One may use the standard formula for the
entropy production of a late-decaying particle
\cite{kolbturner}. This method involves following the
gravitino to photon ratio through all the eras. But since 
here it is assumed that the saxion eventually dominates 
the universe and provides a second reheating at 
roughly $T^{(2)}_{RH} \sim $ 10 MeV, it is more convenient to follow 
the gravitino to saxion number $n_{3/2}/n_S $ 
which remains 
constant.  
Then 
\beq
\left( {n_{3/2} \over n_S} \right)_{T^{(2)}_{RH}} 
= \left({n_{3/2} \over n_S} \right)_{T_{H}} 
=\left( {n_{3/2} \over n_{\gamma}}\right)_{T_H} 
\left({ n_{\gamma} \over n_S}\right) _{T_{H}} ~.
\label{ratio1}
\eeq
>From this expression we find that a more effective 
dilution has a smaller photon to saxion ratio at 
the high temperature $T_H$.
The choice for the initial temperature 
$T_H$ at which to evaluate this expression 
depends on one of two scenarios. 
In the first the saxion begins to oscillate 
during the matter dominated (MD) era of the inflaton. Then 
the particle ratios are evaluated at 
$T_H=T^{(1)}_{RH}$, the initial reheating 
temperature. In the second instance, the 
saxion begins to oscillate during the radiation dominated
(RD) era 
subsequent to the first reheating. Then 
$T_H = T_{osc}$, the temperature at 
which the saxion begins to oscillate.

In the first scenario, 
the saxion is already oscillating at the time of 
the first reheating. Then  
\beq 
\left({n_S \over n_{\gamma}}\right)
_{T^{(1)}_{RH}} = 
{ f^2 \over M^2_{Pl}} \left({ T^{(1)}_{RH} \over m_S} \right) 
{ g_*(T^{(1)}_{RH})  \pi^4 \over 30 \xi (3)  } ~.
\label{ratio2}
\eeq
In addition, the initial value 
of $n_{3/2} / n_{\gamma}$ is approximately
given by $ T^{(1)}_{RH}/M_{Pl}$. 
Putting these two ingredients together 
gives the gravitino to saxion ratio at $T=T^{(1)}_{RH}$. 
This ratio is fixed until the saxion decays, converting
its energy to radiation at a temperature $T^{(2)}_{RH}$.
The desired gravitino to photon ratio is then 
\beq 
\left({n_{3/2} \over n_{\gamma}}\right)_{T=keV} \simeq 
\left({ g_*(keV) \over g_*(T^{(2)}_{RH})} \right)
\left({n_{3/2} \over n_{\gamma}}\right)_{T^{(2)}_{RH} }
=\left({ g_*(keV) \over g_*(T^{(2)}_{RH})} \right)
 \left({n_{3/2} \over n_S}\right) _{T^{(2)}_{RH} } 
\left({n _{S} \over n_{\gamma}}\right) _{T^{(2)}_{RH} } ~.
\eeq
The last ratio is given by 
\beq 
\left({n_{S} \over n_{\gamma}}\right)_{T^{(2)}_{RH}} \simeq 
 \left({ T^{(2)} _{RH} \over m_S} \right) 
{ g_*(T^{(2)}_{RH}) \pi^4 \over 30 \xi (3)   } ~.
\eeq 
Combining these together gives 
\beq 
\left({n_{3/2} \over n_{\gamma} }
\right)_{T=kev} \simeq {M^2_{Pl} \over f^2_a}
{ T^{(2)}_{RH} \over T^{(1)}_{RH}}
\left({g_*(keV) \over g_*(T^{(1)}_{RH})} {T^{(1)}_{RH} \over M_{Pl}}\right) 
= {M^2_{Pl} \over f^2_a}
{ T^{(2)}_{RH} \over T^{(1)}_{RH}} 
\left({n_{3/2} \over n_{\gamma}}\right)_0 
~.
\eeq
Define the dilution factor $\gamma$ to be the double 
ratio of the 
gravitino--photon ratio in  the standard cosmology to 
the gravitino--photon ratio in 
a cosmology with a late decaying saxion. 
That is, 
\beq 
\left({n_{3/2} \over n_{\gamma} }\right)_{T=keV}
 \equiv { 1 \over \gamma} 
\left({n_{3/2} \over n_{\gamma}}\right)_0 
\eeq
where
\beq
\gamma \simeq
{ f^2_a \over M^2_{Pl}}
{T^{(1)}_{RH} \over T^{(2)}_{RH}} ~.
\eeq
This dilution can be substantial, allowing for a higher reheat
temperature. With $f=10^{15}$ GeV, $T^{(2)}_{RH}=10$ MeV, 
$T^{(1)}_{RH}=10^{11}$ GeV
and $m_S =100 $ TeV, 
one finds $\gamma \sim 10^7$.(That the 
mass needs to be specified will become 
clear in the next paragraph). For these parameters 
the gravitino 
abundance is diluted to an allowed amount.


To this point it has been implicitly 
assumed that the saxion begins 
to oscillate during the inflaton MD  era.  
But if the first reheating is high enough then this
assumption 
will no longer be true. More specifically, 
if  
$T^{(1)}_{RH} > 10^{11} \sqrt{m_S/(100 \hbox{ TeV})} $ 
GeV the saxion begins to oscillate during the 
RD era. 
If this occurs, 
then 
this has the important effect of 
reducing the dilution previously estimated. 
To see this, note that in (\ref{ratio1}) we 
need $n_S/n_{\gamma}$ and choose to 
evaluate it at $T_H=T_{osc}$, the temperature the 
saxion begins to oscillate.   
This ratio is given by (\ref{ratio2}), where in that equation
we replace $T^{(1)}_{RH}$ everywhere with 
the lower temperature $T_{osc}$. Then the previous 
estimate of the dilution can be used, provided that
all the high temperature 
particle number ratios are evaluated at $T_{osc}$ 
instead of $T^{(1)}_{RH}$. 
This leads to 
\beq 
\gamma \simeq 
{ f^2_a \over M^2_{Pl}}
{T_{osc} \over T^{(2)}_{RH}} ~.
\eeq
As noted previously, the dilution is in this 
case less effective, but as before, still 
substantial. 
If for instance $T^{(1)}_{RH} =10^{13}$ GeV, 
then without the entropy production from 
the saxion decay $(n_{3/2} /n_{\gamma})_{T=keV} \sim 10^{-7}$ 
is five orders of magnitude too large. 
Now if the 
saxion mass is 10 TeV, it will begin to oscillate 
at $T_{osc} \approx 10^{11}$ GeV. Then with 
$f _a \sim 10^{15}$ GeV and $T^{(2)}_{RH} \sim 
$ 10 MeV, the above result gives $\gamma \approx 
10^7$ which barely provides enough dilution.

There is another point worth emphasizing. The decay constant
appearing in the dilution factor is that of the
{\em decaying} saxion. As we will stress in subsequent
sections, there are cosmological scenarios in
which this particle is not the Saxion. For
the more general saxions
there is no cosmological upper bound on their
decay constant (provided they decay before nucleosynthesis).
The dilution
factor for $f_a \sim M_{Pl}$ is  much larger.

One may also worry about the thermal
production of axinos, or more generally, modulinos.
An estimate for their abundance in the absence
of a decaying modulus can be obtained by comparing with
gravitino production.  The gravitino production rate is proportional
to $1/M_{Pl}^2$; for modulinos it is proportional to $1/f_a^2$.   If the
modulus couples only to gauge fields, the rate is suppressed by an additional
factor of $({\alpha / \pi})^2$.  However, we have already seen that moduli
which couple only in this way are problematic, and we have argued that
we are principally interested in models in which moduli have large tree level
couplings to most standard model matter fields.  So, like gravitino
production, modulino production processes are also enhanced by roughly the
number of light matter fields.  So these processes are quite dangerous.
For $f =10^{15}$ GeV
and $T^{(1)}_{RH}=10^{11}$ GeV, one
finds that subsequent to
the first reheating
${n_{\tilde{A}} / n_{\gamma}} \sim 10^{-1}$.
A further dilution of only $10^7$ is three to four orders of
magnitude too small.
In this case the Saxion cannot provide enough dilution.
But as mentioned above,
if the decaying particle is a saxion with a {\em larger}
decay constant $\sim M_{Pl}$ , then the dilution factor
might be large enough to provide sufficient dilution.

Non--thermal production of gravitinos and modulinos has
been discussed recently in the literature \cite{nontherm,giudice}.
The change in the inflaton potential near the end of its
life can result in a rapid change in the fermion masses,
leading to production of these particles. More generally,
there will be particle production for particles
that are not conformally coupled. Whether
this leads to an overproduction of particles depends on
the details of the inflationary model. But reheat
temperatures as low as $10^{2}$ GeV may be required \cite{nontherm}.
For example, the dangerous relic to
photon ratio can be as large
as $n_X/n_{\gamma} \sim \beta T_{RH}/M$, where
$M \sim 10^{15}$ GeV characterises the change in mass scales at the end
of inflation, and $\beta$ is a number that depends on the
dependence of the modulino mass on the inflaton \cite{nontherm}.
A large value of $n_X/n_{\gamma} \sim 10^{-5}$
$(T_{RH}=10^{11}$ GeV and $M=10^{15}$ GeV, $\beta \sim 10^{-1})$
is then not unreasonable.
In the conventional
scenario this is much too large. But we
have seen that the late decay of either the
Saxion or another saxion with a larger decay constant,
can produce a dilution of $10^7$ or much larger.
This amount of dilution might be sufficient.

Since coherent overproduction 
occurs for fermions that have a rapidly changing mass 
during the end of inflation, it is clear that the issue is whether
there are any fermions that are coupled to the inflaton 
\cite{nilles}. 
Thus in a model where the susy breaking sector 
today is decoupled (except by supergravity) from the inflaton sector, 
today's gravitino would not have been coherently overproduced. 
But the gravitino 
then - the inflatino - would have been. This isn't 
necessarily problematic, since the inflatino 
has a very short lifetime. But if other modulinos 
${\cal M}$ are present, then 
during 
inflation they may have acquired a 
mass $m \sim H$ from the Kahler potential operators 
\beq 
{I^{\dagger} I \over M^2_{Pl}} {\cal M} {\cal M} ~, 
{I^{\dagger} \over M_{Pl}} {\cal M} {\cal M} ~,
\eeq
or, acquired mass mixing with the inflatino, 
of the same order, through superpotential interactions \cite{giudice}. 
Depending on the model, 
it may have been overproduced during the transition 
to the MD inflaton era \cite{giudice}. 
If the modulinos were 
overproduced and their late decays are problematic, then the 
dilution provided by a late decaying 
scalar might be welcome.  

In sum, modulino
production is a significant constraint on the picture we have presented here.
The decays of a massive modulus with a large
decay constant provide 
the simplest way to sufficiently dilute these particles.

\section{Cosmological Limits on the QCD Axion}

In this section, we focus on the general question of the limits on
the QCD Axion in supersymmetric theories.  There are many possibilities
we can consider, both for models of supersymmetry breaking and
for the behavior of the early universe.  We can consider
supersymmetry breaking at intermediate scales (as we have up to now
for generic axions) or
gauge mediation, for example.
We will focus in this section, as we have
up to now, on supersymmetry breaking
at intermediate scales, saving the discussion of low energy
breaking for section 6.

Again, we argue that it only makes senses to
consider limits on the Axion in the context of acceptable Saxion
cosmologies.  We will see that there are cosmological windows on the
Axion beyond the usual ones.

Of the many possible cosmic histories we might consider, we will
focus on a limited set.
First,  there might be other moduli, with decay
constants larger than $f_A$, which dominate the energy density at least
from the time that $H=m_{3/2}$.  Alternatively, we might imagine
that there are no such moduli, and that the universe reheats after
inflation to some high temperature.  In that case we ask if and
when the Saxion comes to dominate the energy density.

\subsection{Conventional Axion Cosmology}

In the conventional cosmology the Axion does not begin to
oscillate until the temperature is around a few GeV. The
actual value depends on the decay constant. A rough
estimate for when the Axion begins to oscillate
may be obtained using
$m_a(T) \simeq 3 H(T)$, where
the temperature-dependent
mass is \cite{turner}
\beq
m_a(T) = 0.1 m_a(T=0) \times \left({\Lambda_{QCD} \over T }\right)^{3.7} ~.
\label{axionmass}
\eeq
This expression
is valid for $\pi T \gg \Lambda_{QCD}$, where here
$\Lambda_{QCD} \equiv 200$ MeV. For lower temperatures
the Axion mass is to a good approximation given by its zero
temperature value.
Since the Axion potential is temperature dependent, the energy
density in Axions does not redshift as $R^{-3}$, but instead as
\beq
\rho_a(T)= {m_a(T) \over m_a(T_0)} {R^3_0 \over R^3} \rho_a(T_0).
\label{axionenergy}
\eeq
In the conventional picture, the
Axion begins to oscillate at a temperature much larger
than the QCD scale.  Requiring that the energy in Axions constitutes
less than 1/3 of the current energy density
gives the usual limit, $f_a \ltap
 3\times 10^{11}$ GeV.

\subsection{Axion Evolution in the Presence of a Saxion}

Suppose that supersymmetry is broken at an intermediate scale as
in supergravity models.  In this case, the Saxion mass will be of
order TeV (perhaps tens of TeV).  As a result, the conventional assumption
of radiation domination at the QCD scale is not necessarily
correct. We will first
suppose that the Axion supermultiplet is the only
moduli supermultiplet.  This scenario has been
discussed in \cite{liftingdecayconst} \cite{saxioncosmo} , so
most of this subsection is review.



In the conventional inflationary scenario the
Saxion begins to oscillate during the matter-dominated
phase of the inflaton.
Assume, as is conventional,
that all of the energy of the inflaton is transferred to radiation
during reheating.
Then the fraction of energy stored in
the Saxions immediately at the start of the first
radiation-dominated era is $
\rho_s \simeq f^2_A \rho_R /  M^2_{Pl}~.$
The Saxions will come to dominate the universe when the
temperature has dropped to
$T_s \simeq ( f^2 _A /M^2_{Pl}) T^{(1)}_{RH} ~.$
Here $T^{(1)}_{RH}$ is the first reheating
temperature.
In order for the late decays of Saxions
to be effective, this cross-over temperature must, first,
occur before the saxion decays.  Otherwise, Saxion decays will
not effect the Axion cosmology appreciably.  This gives a
{\em lower bound} :
\beq
f_A > {M_{Pl}^{5/6} m_{s}^{1/2} \over (T^{(1)}_{RH})^{1/3}}
=10^{13.5} ~\hbox{GeV}~ \left({m_{s} \over \hbox{TeV}}\right)^{1/2}
\left({10^{9} ~\hbox{GeV} \over T^{(1)}_{RH} } \right)^{1/3}
\eeq
(neglecting factors of order one).
If this bound is not satisfied, the Axion will begin
to oscillate when the universe is radiation dominated,
and the conventional limits apply. This bound may instead
be viewed as requiring that the initial reheat
temperature
must be larger than
\beq
T^{(1)}_{RH} \gtap 3 \times 10^4 ~\hbox{GeV} ~ \left(
{10^{15} ~\hbox{GeV} \over f_A} \right)^3 \left({m_s \over
\hbox{TeV}}\right)^{3/2}~.
\label{saxiont}
\eeq

There is in addition another requirement,
namely that
the cross-over temperature must be above a GeV, so
that the Axion begins to oscillate during the
Saxion-dominated phase.
This requires
\beq
f_A \gtap 10^{13.5} \hbox{GeV} \left({10^{9} \hbox{GeV}
\over T^{(1)}_{RH}} \right)^{1/2}
\label{GeVbound}
~.
\eeq
This bound is independent of the Saxion mass (assuming
it begins oscillating during the end of inflation).
Clearly a high reheat temperature is needed.

The reason for this second bound may be understood as follows.
The relevant quantity is the
Axion energy density at the time of Saxion decay. Due
to the temperature dependence of the potential,
the energy density increases adiabatically relative
to the $R^{-3}$ redshift by an amount $m_A(T)/m_A(T_{osc})$.
Inspecting (\ref{axionmass}), this factor can be
large. This is as in the conventional cosmology, precisely because
the Axion begins to oscillate when the temperature
is around a GeV.

But if the Saxion (or any other modulus) begins
to dominate the energy before the universe cools
to a GeV, then the story is different.
The reason is that
in a matter dominated universe the Hubble parameter must,
by definition, be
much larger than its value in a radiation dominated universe
at the same temperature. When the temperature
reaches a GeV in the Saxion-dominated universe,
the Hubble parameter is still too
large and the Axion hasn't begun to oscillate.
Instead it begins
to oscillate when the temperature is roughly a 100 MeV.
At these
temperatures the Axion mass is approximately
given by its zero temperature
value and there is no large enhancement.

Assuming the Saxion decays solely
into radiation,
the ratio of energy in Axions
to radiation at the start of the second radiation era is
$f^2_A/M^2_{Pl}$, dropping factors of order one.
Requiring that the current Axion energy density be less than
about a third of the critical energy density gives the
bound
\beq
f_A \ltap 10^{15} \sqrt{\left({m_a(T^M_{osc}) \over
m_a(0)}\right)}
{h_0 \over 0.7} \sqrt{10 \hbox{MeV}
\over T^{(2)}_{RH}} ~ \hbox{GeV} ~.
\eeq

But stated in this way the upper bound is a little misleading, since
not all values of the decay constant up to
$10^{15}$ GeV are allowed. As previously indicated,
there are {\em lower bounds}.
In fact as we can see, there is a
small window between $\sim 10^{13.5}$ GeV and
$\sim 10^{15}$ GeV in which there are no cosmological
problems.
For smaller values of the Axion
decay constant there is no Saxion cosmological
problem. But there is a conventional
Axion abundance problem since the Saxion
decays are not useful.
For Saxion decay constants below about
$\sim 3 \times 10^{11}$ GeV,
the conventional analysis applies and there is neither
a Saxion nor an Axion cosmological problem.

Finally, one also requires that the second reheating
from the Saxion decay is above an Mev, but below 100 MeV.
This does limit the parameter space. For instance,
these lower and upper bounds imply
$10^{16.5}$ GeV $\gtap f_A \gtap 10^{14.5}$ GeV for
a Saxion mass of order TeV (neglecting factors of order 1),
and scale as $m_S^{3/2}/$TeV$^{3/2}$.


\subsection{Axion Evolution in the Presence of Another Modulus}


We saw in the previous section that
an Axion decay constant in an intermediate
range was not allowed if the Axion is the only modulus.
If there is a second modulus with a
different decay constant the limitations of the single-modulus
scenario can be relaxed.
By having a larger decay constant for the new modulus,
two things happen, which both favor allowing the full
window of Axion decay constants : first, it is easier
for the new
modulus to dominate the universe before it
reaches a GeV, and second, the requirement
that the second reheating temperature is not too high is
easier to satisfy.  We now show
that this scenario
allows the window $f_A < 10^{15}$ GeV with no lower
cosmological bound.

Suppose that shortly after inflation, the universe is dominated
by a massive modulus,
other than the Saxion. This will
occur if its decay constant is
$F \sim M_{Pl}$. It is important
that the decay constant of this
other modulus is unrelated to the Saxion and Axion decay
constant $f_A$.
In this case, the Saxion also begins to oscillate
essentially immediately.  It carries a fraction of the energy
density of order $f_A^2/M^2_{Pl} \sim f^2_A/F^2$.
If $f_A \ll F$, the Saxion carries
only a small fraction of the energy density of the universe, and
decays long before the massive modulus.
The modulus is assumed sufficiently heavy (greater than 20 TeV)
that it reheats the universe above nucleosynthesis
temperatures, but below the QCD scale.
At this time, the energy density in Axions is
suppressed relative to that in moduli by a factor $f_A^2/M_{Pl}^2$,
which is sufficiently small that Axions don't dominate before
recombination if $f_A < 10^{15}$ GeV.  So this picture is
self-consistent.
For Axion decay constants less than $10^{15}$
GeV, there is neither a Saxion nor an Axion density problem.
Note that the reheating in the modulus decay also dilutes
gravitinos which may have been produced during the first reheating.

For smaller values of the modulus decay constant
$F$ the subsequent cosmology
and relic Axion abundance
in this scenario is determined by several more parameters:
the Axion decay constant $f_A$;
the reheat temperature $T^{(1)}_{RH}$
due to the inflaton decay; and
the reheat temperature $T^{(2)}_{RH}$ due to the decay of
the scalar modulus.  We have explored this parameter space sufficiently
to establish that there is an
%
additional small window of
allowed axion decay constants.

\section{Gauge Mediation}

So far, we have assumed that supersymmetry is broken at an
intermediate scale, and the gravitino mass is of order $1 ~{\rm
TeV}$.   An alternative possibility is that supersymmetry is
broken at a lower scale, as in gauge mediation.

It has been argued that gauge mediation is likely to arise in
theories where there are no moduli, or where the moduli are fixed
by supersymmetry-preserving dynamics at a very high energy scale.
In this case, all of the moduli are much more massive
than the TeV scale, and there is no moduli problem.  However, there are also no
axions to solve the strong CP problem.  In the context of gauge
mediation, there are viable alternative solutions to the strong
CP problem; these
will be discussed in the next section.

If there are moduli, then in gauge-mediated theories they
are relatively -- possibly
very -- light.  To get some feeling for the issues, suppose
that the scale of supersymmetry breaking is of order $\sqrt{F}$.
As we have discussed, the saxion mass in this case is likely to be
of order $\epsilon^{1/2} {F \over f_a}$, where $\epsilon$ is typically
some combination of loop factors.
The leading saxion
coupling to Standard Model matter is
given by the axion-like interactions with
gauge fields, induced by integrating out heavy vector-like
fields. This has a suppression factor of $\alpha/2 \pi$, so
the saxion lifetime is then of
order:
\beq
\Gamma \approx {1 \over 2 \pi}
\left({\alpha_s \over 2 \pi}\right)^2  {\epsilon^{3 /2} F^{3} \over f_A^5}
\eeq
Correspondingly, the reheat temperature is of order
\beq
T_R \approx {(10^{-2} \epsilon^{3/4} F^{3/2}M^{1/2}_{Pl}) \over f_A^{5/2}}
= 10 ~\hbox{GeV}~ \left({\epsilon \over 0.2}\right)^{3/4}
\left({\sqrt{F} \over 10^8 \hbox{GeV}}\right)^{3}
\left({10^{12} \hbox{GeV} \over f_A}\right)^{5/2}.
\eeq
In other words, if $f_A$ is $3 \times 10^{11}$ GeV, it is necessary that the
scale of supersymmetry breaking be greater than about $10^{6.5}$ GeV in order
to reheat to above 10 MeV.
For $f_a=10^{15}$ GeV, the scale of supersymmetry breaking must be of
order $10^{9.5}$ GeV or higher 
to obtain sufficient reheating.  From the
point of view of gauge mediation, such a high 
scale for
supersymmetry breaking might be problematic: flavor violating
soft masses from  
 Planck suppressed 
terms are no longer negligible.

Observe that
the reheat
temperature here scales as $f^{-5/2}$, whereas in gravity
mediation it is much softer, $\sim f^{-3/2}$.
Achieving a sufficient reheating  
is much more difficult in gauge mediation. This will 
be important to what follows. 

For a low scale of supersymmetry breaking, 
$\sqrt{F} \sim 10^{5}$ GeV$-3\times 10^{6}$ GeV, requiring 
sufficient reheating implies that the Axion decay 
constant cannot be too large, $f_A \ltap 10^{10}$ GeV$-3 \times 10^{11}$ GeV. 
In this range
both the Saxion and Axion problems are 
solved and the Axion could be the dark matter candidate. 
But for larger values of the 
Axion decay constant, $f_A \geq 3 \times 10^{11}$ GeV, 
the Saxion problem still exists since it 
decays too late. 

In cases where the saxion problem is solved, the axion problem is
often solved as well.  As before, we need to ask at what
temperature the saxion comes to dominate the energy density of the
universe.  Consider the case $f_A = 10^{15}$ GeV, $\sqrt{F} =
10^{10}$ GeV. 
In this case, the saxion mass is of order $
\sqrt{\epsilon} \times 100$ TeV,
and the estimates are similar to those we encountered in the
previous section.  In particular, both the saxion and axion
problems are solved. But as already mentioned, 
such a high scale may be problematic for 
flavor violating processes. 
As we decrease $\sqrt{F}$, we need also to
decrease $f_A$.

\section{Alternative Solutions to the Strong CP Problem}

We have seen that there are situations where an 
 Axion solution to the Strong CP problem is
likely to be unworkable.  For
example, for gauge mediation, with a SUSY breaking
scale below $3 \times 10^6$
GeV and $f \gtap 3 \times 10^{11}$ GeV, 
the Saxion problem is not easily solved.  
Given that, on other grounds, such low scale
supersymmetry breaking is a plausible picture of how nature works,
it is interesting to examine carefully other solutions of the
strong CP problem \cite{nelson-barr,schmaltz}.
We have already commented on the possibility
of a massless $u$ quark.  In this section we briefly comment on
the Nelson-Barr mechanism \cite{nelson-barr}.
The points we will make have largely
appeared earlier in \cite{dkl}, and more recently in
\cite{schmaltz}.

In \cite{dkl}, it was shown that implementing the Nelson-Barr
mechanism is, in general, rather difficult in supersymmetric
theories.  The problem is that loop corrections involving squarks
and gauginos give large corrections to $\theta$ unless there is a
very high degree of degeneracy.  In \cite{schmaltz}, it was noted
that gauge mediation can provide the needed level of degeneracy.
So low--energy 
gauge mediation, which is precisely the case where one might need an
alternative to axions, is a situation in which there is an
alternative solution to the strong CP problem.

\section{Conclusions:  Detectable Axion Dark Matter}

We have seen that in supersymmetric theories, the problems of
Saxion cosmology are much more serious than those associated with
Axions.  Mechanisms which solve the saxion problem usually modify
the limits on the axion decay constant.  One is thus left with
several possibilities:
\begin{itemize}
\item  There is no Axion.  The strong CP problem is solved by a
massless $u$ quark, or by a variation on the Nelson-Barr
mechanism.  We saw that this view is almost inevitable if one has
gauge mediation, with a supersymmetry breaking scale below $3 \times 10^6$ GeV 
and a large Axion decay constant. 
\item  The Axion solves the strong CP problem, but does not
constitute the dark matter; it's decay constant is, say, $10^{15}$
GeV, a scale which might plausibly emerge from string theory.
Such a picture emerges naturally in the case of gravity mediation.
\item  The Axion solves the strong CP problem, constitutes the
dark matter, but is not detectable in forseeable experiments
because of its large decay constant.  Again, this can readily
emerge from gravity mediation.
\item  The Axion solves the strong CP problem, constitutes the
dark matter, and its decay constant is such that it can be
detected in future experiments.
\end{itemize}

While the discussion of this paper makes clear that the last possibility
is far from inevitable, it also suggests that it might be possible.
Indeed, our simple field theory model for axions suggests that
this could come about naturally.  If the symmetries of the theory
are such that $n=1$, then $f_A \approx 10^{11}$ GeV.  The axion,
in such a model, will be sufficiently light provided, for example,
one has a discrete symmetry which insures that the leading
PQ violating correction to $W$ goes as $S^{9}$ or larger.

Indeed, the usual axion limit poses the question:  where does the
scale $10^{11}$ GeV come from?  From the point of view of
supersymmetry, the mechanism we described above was always a
natural candidate.  In contrast, in non-supersymmetric theories,
this extra scale must simply be postulated, and raises all of the
usual questions of hierarchy.  Having added this scale to
the theory, one has introduced a new fine tuning problem for the Higgs
which is far worse than that connected with the strong CP problem.

\noindent
{\bf Acknowledgements:}

\noindent
We thank Ed Witten for asking a set of questions which prompted
this investigation, and Scott Thomas for discussion of a number
of issues, particularly saxion decays to axions.
This work is supported in part by the U.S.
Department of Energy. MG would like to thank the Aspen
Center of Physics where part of this work was completed.


\end{document}